\documentclass[lettersize,journal]{IEEEtran}
\usepackage{amsmath,amsfonts}
\usepackage{algorithmic}
\usepackage{algorithm}
\usepackage{array}
\usepackage[caption=false,font=normalsize,labelfont=sf,textfont=sf]{subfig}
\usepackage{textcomp}
\usepackage{stfloats}
\usepackage{url}
\usepackage{verbatim}
\usepackage{graphicx}
\usepackage{cite}
\usepackage{multirow}
\usepackage{threeparttable}
\usepackage{xcolor}
\begin{document}

\title{Confidence analysis-based hybrid heartbeat detection for ballistocardiogram using template matching and deep learning}

\author{Dongli Cai, Xihe Chen, Yaosheng Chen, Hong Xian, Baoxian Yu, Han Zhang,~\IEEEmembership{Member,~IEEE}
\thanks{Dongli Cai, Xihe Chen, Yaosheng Chen, Baoxian Yu and Han Zhang are with School of Electronic Science and Engineering (School of Microelectronics), South China Normal University, Foshan 528200, China and also with Guangdong Provincial Research Center for Cardiovascular Individual Medical and Big Data, South China Normal University, Foshan 528200, China (e-mail: zhanghan@scnu.edu.cn; yubx@m.scnu.edu.cn).}
\thanks{Hong Xian is with West China Hospital, Sichuan University, Chengdu 610044, China.}}


\IEEEpubid{0000--0000/00\$00.00~\copyright~2021 IEEE}
\IEEEpubidadjcol 

\maketitle

\begin{abstract}

Heartbeat interval can be detected from ballistocardiogram (BCG) signals in a non-contact manner. Conventional methods achieved heartbeat detection from different perspectives, where template matching (TM) and deep learning (DL) were based on the similarity of neighboring heartbeat episodes and robust spatio-temporal characteristics, respectively, and thus, performed varied from case to case. Inspired by the above facts, we propose confidence analysis-based hybrid heartbeat detection using both TM and DL, and further explore the advantages of both methods in various  scenarios. To be specific, the confidence of the heartbeat detection results was evaluated by the consistency of signal morphology and the variability of the detected heartbeat intervals, which could be formulated by the averaged correlation between each heartbeat episode and the detected template and the normalized standard deviation among detected heartbeat intervals, respectively, where the results with higher confidence were remained. In order to validate the effectiveness of the proposed hybrid method, we conducted experiments using practical clinical BCG dataset with 34 subjects including 924,235 heartbeats. Numerical results showed that the proposed hybrid method achieved an average absolute interval error of 20.73 ms, yielding a reduction of 29.28 ms and 10.13 ms compared to solo TM and DL methods, respectively. Besides, case study showed the robustness of heartbeat detection of TM and DL to individual differences and signal quality, respectively, and in turn, validated that the hybrid method could benefit from the complementary advantages of both methods, which demonstrated the superiority of the proposed hybrid method in practical BCG monitoring scenarios.

\end{abstract}


\section{Introduction}
\IEEEPARstart{L}{ong-term} monitoring of heart rate variability is of great significance for the early detection of cardiovascular disease (CVD) and emotion analysis \cite{rajendra2006heart},\cite{zhu2019heart},\cite{shokouhmand2021mean}. Electrocardiogram (ECG) has been regarded as the thumb rule for cardiac monitoring in clinical scenarios \cite{drew2004practice},\cite{sandau2017update},\cite{kotalczyk20212020}. However, the direct attachment between electrodes and skin may cause discomfort during long-term nocturnal monitoring \cite{10288217},\cite{10449345},\cite{10551856}. Ballistocardiogram (BCG) has been regarded as a promising alternative not only owing to the great consistency in heartbeat intervals with ECG, but also owing to its non-contact acquisition manner\cite{liu2021motion},\cite{10387261},\cite{yu2025proof}. Particularly, BCG sensors can be easily embedded in a mattress or bed thanks to the advanced sensing techniques \cite{9103629},\cite{escobedo2024bed},\cite{10835744}, which provide promising long-term nocturnal monitoring solutions. Based on reliable detection, BCG can be further applied to diagnose atrial fibrillation \cite{jiang2022topological},\cite{su2022atrial},\cite{11018225} and heart failure \cite{aydemir2019classification},\cite{feng2023machine},\cite{zhan2024non}.

\begin{figure}[tbp]
\centering  
\includegraphics[scale=0.37]{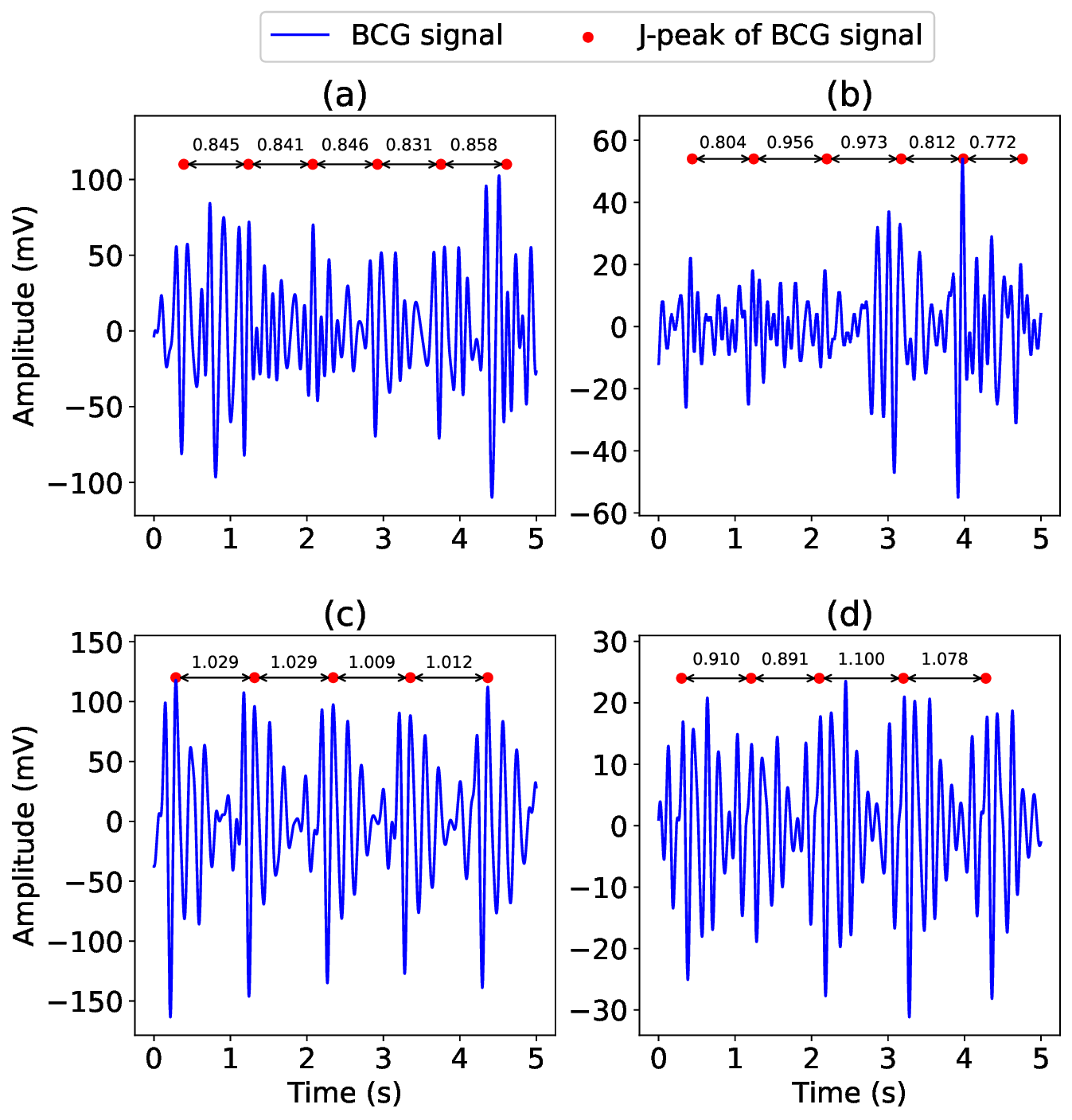}
\caption{5 s BCG signals recorded from different subjects. (a)-(b), for the case of BCG signals in low signal-to-noise ratio scenarios, (a) sinus rhythm; (b) arrhythmia.  (c)-(d), for the case of high signal-to-noise ratio, (c) sinus rhythm; (d) arrhythmia.}  
\label{Figure 2} 
\end{figure}

\IEEEpubidadjcol
Similar to ECG, heartbeat intervals in BCG signals can be derived from the intervals between neighboring J-peaks. However, the complexes of BCG are much more complicated than ECG, and thus, numerous methods were proposed to address the above issue, where template matching (TM) and deep learning (DL) methods have been regarded as the most representative ones \cite{yu2025proof}. To be specific, TM employs the dynamical templates to detect J peaks by evaluating the correlation between BCG signals and templates. Therefore, the performance of TM is highly determined by the robustness of the constructed templates. The template modeling can be achieved by ensemble averaging \cite{shin2008automatic},\cite{nagura2018estimation}, clustering \cite{paalasmaa2014adaptive},\cite{xie2019personalized}. Nevertheless, the robustness of template is mainly determined by the quality of original BCG signals, which is uncontrolled in practical scenarios. As shown in Fig. \ref{Figure 2}(a) and (b), template modeling may fail due to the insignificant similarity among BCG complexes of each heartbeat, which may raise incorrect heartbeat detections.

DL-based heartbeat interval detection methods rely on intrinsic morphological features or time sequence of BCG complexes rather than the consistency among neighboring signals. Therefore, convolutional neural network and  recurrent neural network (including their variants, e.g.,  UNet and gated recurrent unit) have been widely employed to extract spatio-temporal features\cite{jiao2021non},\cite{mai2022non},\cite{10810416}. However, BCG signals may suffer from individual differences, such as insignificant J-peaks raised by unexpected measurement postures and arrhythmia, as shown in Fig. \ref{Figure 2} (c) and (d), respectively. The above facts may raise challenges in generalization of DL models for BCG heartbeat interval detection in practical scenarios.

Inspired by the pros and cons of TM and DL methods, we propose a confidence analysis-based hybrid heartbeat detection method for BCG signals using representative TM algorithm and DL model to complement each other, and validate its superiority with practical clinical dataset. The contributions of this paper are summarized as follows.

\begin{itemize}
    \item[$\bullet$] We propose a confidence analysis-based hybrid heartbeat detection method for BCG signals using TM and DL. Specifically, the confidence analysis w.r.t. the morphological and rhythmic constraints on BCG signals is utilized to evaluate the acceptability of heartbeat detection results performed by specific TM algorithm or DL model, and in turn, to hybridize the most reliable detected heartbeats of a higher confidence level from the acceptable candidates detected using either one.
    \item[$\bullet$] To validate the effectiveness of the proposed method, we constructed a practical BCG dataset by recruiting 34 subjects, with 924,235 heartbeats recorded during nighttime sleep scenarios. Numerical results showed that the proposed method can improve the performance in terms of the average absolute interval error by 10.13 ms and 29.28 ms in comparison with the representative DL and TM, respectively, which demonstrated the superiority of the hybrid heartbeat detection compared to individual methods.
    \item[$\bullet$] Through comprehensive analysis, for the first time, we demonstrate the potential that TM algorithm and DL model, to some extent, can complement each other. To be specific, TM can benefit from the robustness of the template extracted through the time-series BCG signals, and can better effectively adapt to individual differences in high signal quality scenarios in comparison with DL, while DL model can extract the contextual information of signals, and has relatively superior generalization ability in low-quality signals to TM. The above discoveries further demonstrate the significance of the proposed hybrid detection method against individual differences by enjoying complementary advantages of TM- and DL-based heartbeat detection in practical BCG monitoring scenarios.
\end{itemize}

The rest of the paper is organized as follows. Section \ref{S2} presents the method of the proposed confidence analysis-based hybrid heartbeat detection for BCG signals. Experimental setup and numerical results are shown in Sections \ref{S3}. Finally, Section \ref{S4} concludes this paper.

\section{METHOD}
\label{S2}

\subsection{Measurement Protocol and Data Acquisition}
\begin{figure}[tbp]
\centering  
\includegraphics[scale=0.13]{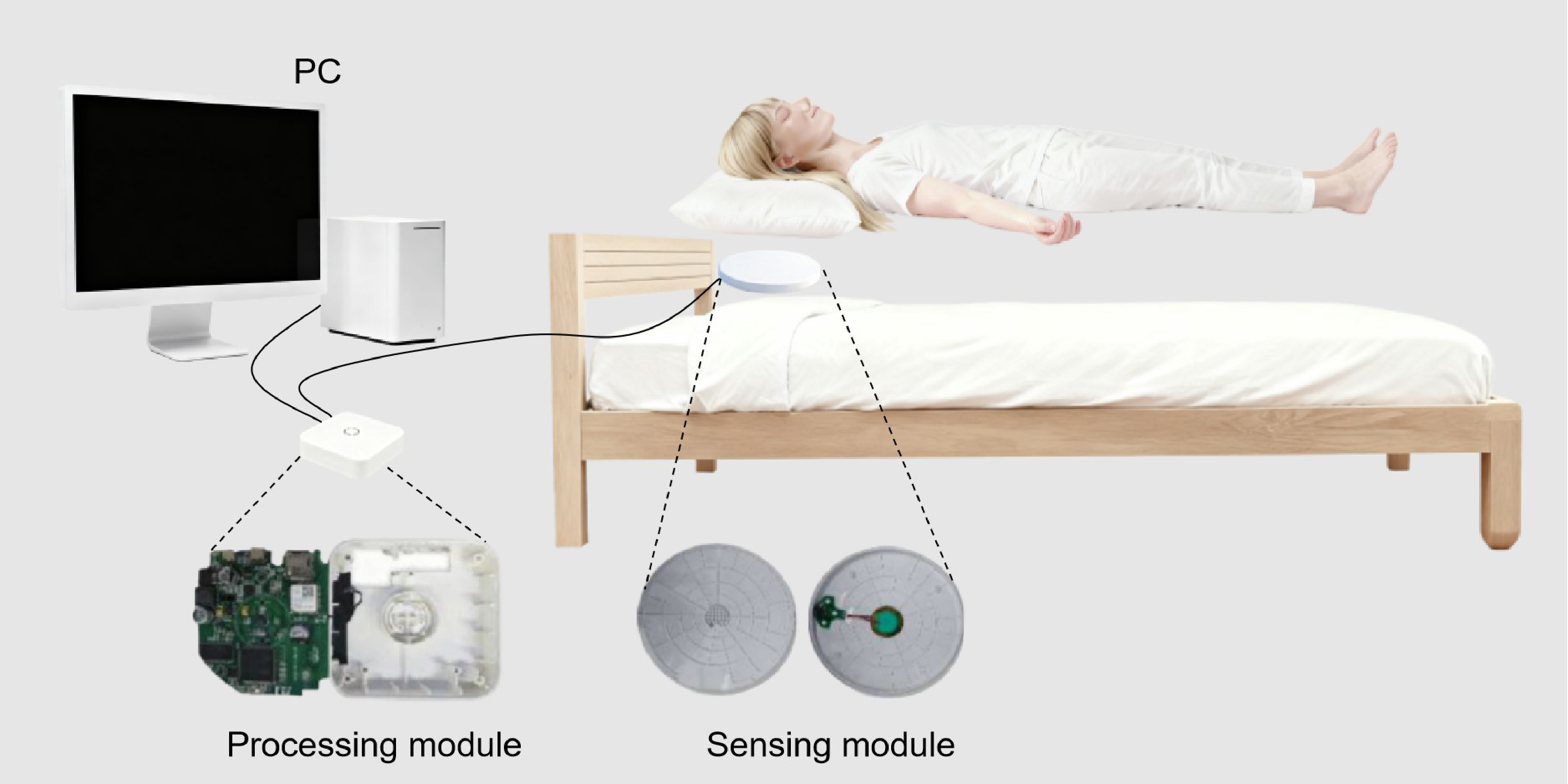}
\caption{Non-contact cardiac monitoring system.}  
\label{Signal acquisition} 
\end{figure}

\begin{figure}[tbp]
\centering  
\includegraphics[scale=0.1]{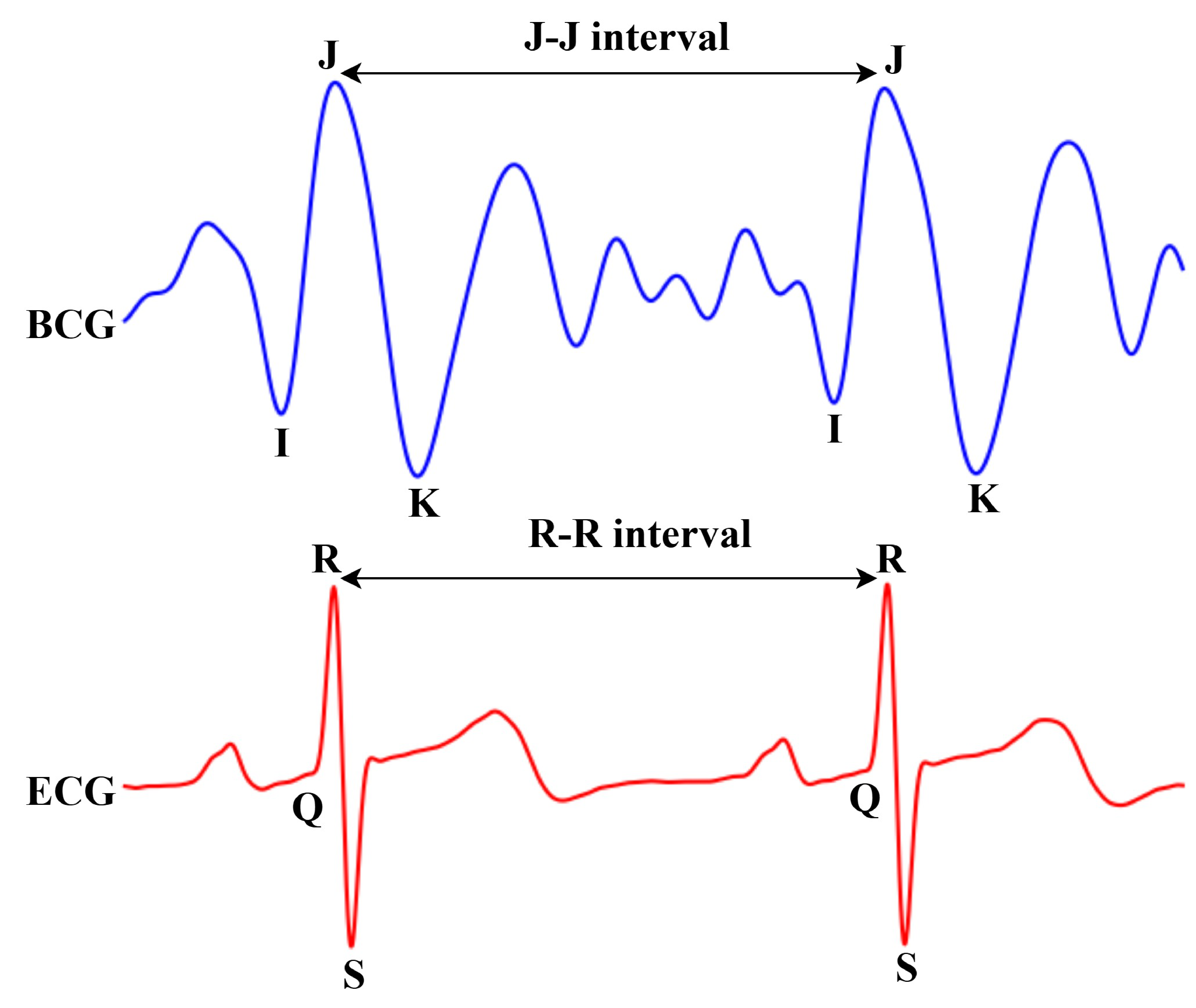}
\caption{Consistency of Heartbeat Intervals Between BCG and ECG.}  
\label{BCGECG} 
\end{figure}

The study was approved by the Academic Committee of South China Normal University (Approval No.: SCNU-PHY-2024-015). The non-contact cardiac monitoring system used in this study is presented in Fig. \ref{Signal acquisition}, which comprises a sensing module and a processing module. To elaborate a little further, the sensing module consists of a thin circular metal plate embedded with a suspended piezoelectric ceramic sensor (KD-35B-26ENW35C-SN), composed of a 25 mm diameter ceramic disc and a 35 mm diameter metal plate with a combined thickness of 45 mm. Based on the sensor placed under the pillow, the vibrations of cardiac ejection, respiratory effort (possibly mixed with artifact motion) were recorded. Then the processing module (a microcontroller STM32F446ZET6) converts the analogy signals into digital signals and realizes the signal acquisition. After band-pass filtering, the BCG signals can be obtained. As the ground truth, ECG signals were simultaneously recorded during BCG acquisition by using Alice 5 PSG with the same sampling frequency to that of BCG (1 kHz). After signal acquisition, preprocessing is performed to the recorded BCG signals. Specifically, a third-order Butterworth band-pass filter with a frequency band ranging from 1-10 Hz is used to suppress the effect of respiratory effort and noise before signal detection\cite{hai2020heartbeat}. In BCG, the heartbeat interval is commonly quantified using the J–J interval,  which serves as a functional analogue to the R–R interval in ECG, as depicted in Fig. \ref{BCGECG}.

The experimental data were recorded from 34 subjects with sleep apnea syndrome (recruited from the First Affiliate Hospital of Guangzhou Medical University), including 31 males and 3 females (age: 46.79 ± 12.83, BMI: 26.01 ± 3.26). 11 out of 34 subjects had arrhythmia, and the rest exhibited sinus rhythm. Data for all subjects were recorded during sleep in real world, and the total duration of data recording is 16605 minutes, including 924,235 heartbeats. In the experimental dataset, the recorded average heart rate ranges from 46.64 to 79.71 bpm, with the lowest and highest instantaneous heart rate of 36.52 and 145.81 bpm, respectively.

\subsection{Confidence Criteria}

Next, we employ two widely used heartbeat detection methods: TM and DL. In order to derive more precise and reliable heartbeat intervals from BCG signals, we propose a confidence analysis-based hybrid heartbeat detection method to enjoy the complementary advantages of both TM and DL. To be specific, we propose confidence criteria by taking the morphological and rhythmic consistency into consideration, to determine the acceptability of the heartbeat detection results performed by specific algorithm or model. Based on confidence analysis, hybrid detection is proposed to hybridize the more reliable detected heartbeats from the acceptable candidates using both the representative TM and DL.

\begin{figure}[tbp]
\centering  
\includegraphics[scale=0.25]{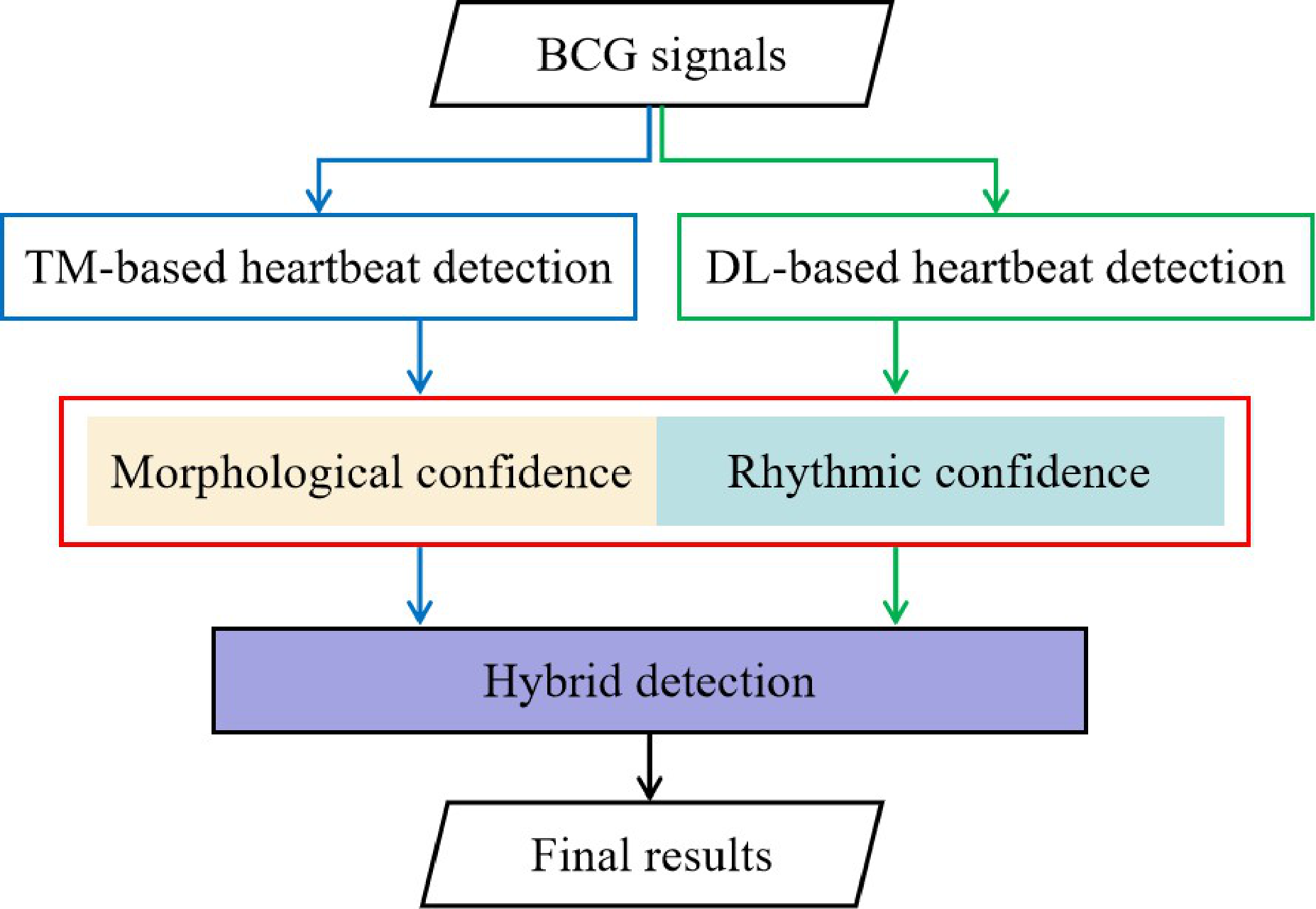} 
\caption{Procedures of the proposed confidence criteria.}  
\label{Fig:Con Cri} 
\end{figure}

We first exclude the unacceptable heart rates that are not between 30 to 180 beats per minute (bpm). Subsequently, we design the confidence of the detected BCG signals w.r.t. the consistency in terms of the morphology and rhythm among time-series consecutive BCG signals. In principle, the confidence criteria are proposed to determine the acceptability of heartbeat detection for BCG signals, where the overall procedures are shown in Fig. \ref{Fig:Con Cri}. 

\subsubsection{Morphological Confidence (MC)}
Motivated by the similarity among neighboring complexes of each heartbeat \cite{6862843}, we consider the morphological confidence index $c_1$ based on the mean of the normalized correlation coefficient between the detected BCG complexes $\mathbf{s}_i$ and the corresponding BCG template $\overline{\mathbf{s}}$. The morphological confidence evaluating the complex consistency of the detected results using either TM or DL, is designed as
\begin{equation}
\label{eq:c1}
c_1 = \frac{1}{M} \sum_{i=1}^{M} {
\frac{Cov(\mathbf{s}_i, \overline{\mathbf{s}})}
{\sqrt{Cov(\mathbf{s}_i, \mathbf{s}_i) \times Cov( \overline{\mathbf{s}}, \overline{\mathbf{s}})}}}
\end{equation}
where $M$ denotes the number of detected heartbeats, also known as J-peaks. $Cov(\cdot, \cdot)$ denotes the covariance of two sequences. $\mathbf{s}_i$ and $\overline{\mathbf{s}}$ are the initially detected BCG signal vectors of the $i$-th heartbeat and the modeled template by consecutive BCG signals, respectively. We define $T_\mathrm{MC}$ as the confidence threshold of $c_1$. To elaborate a little further, as regards the threshold $T_\mathrm{MC}$, only when the index of MC is higher than the threshold, i.e., $c_1 \geq T_\mathrm{MC}$, the detected results using either TM or DL can be considered reliable, otherwise, the detected results will be discarded. According to (\ref{eq:c1}), $c_1$ ranges from 0 to 1, where a lager value of $c_1$ indicates higher similarity among the neighboring BCG signals within a specific epoch. As $c_1 \to 1$, clearly, higher signal-to-noise ratio of the corresponding epoch is expected, and the level of confidence of the detected results becomes higher as well. 

\subsubsection{Rhythmic Confidence (RC)}
Motivated by the pseudo periodicity of the inter-beat intervals of BCG signals, we consider the rhythmic confidence criterion by taking inter-beat variability into consideration. In short-time BCG episode, the heartbeat intervals changes almost quasi-stationary, especially in resting scenarios. Therefore, we consider the rhythmic confidence by using the normalized standard deviation of N-N intervals (SDNN) to evaluate the inter-beat variability within a BCG episode \cite{hayn2012qrs},\cite{8253824},\cite{9452120}. To be specific, the rhythmic confidence $c_2$ can be defined as
\begin{equation}
\label{eq:c2}
c_2 = \frac{1}{\overline{d}}\sqrt{\frac{\sum_{i=1}^{M-1}{(d_i-\overline{d})^2}}{M-2}} 
\end{equation}
where $d_i$ denotes the $i$-th heartbeat interval determined by using either TM- or DL-based method, $\overline{d}$ denotes the average heartbeat interval. Let $T_\mathrm{RC}$ denote the threshold level of RC. To be specific, any detected results $c_2>T_\mathrm{RC}$ are considered unreliable and will be discarded. We can learn from (\ref{eq:c2}) that a smaller value of $c_2$ indicates a heartbeat consistency over consecutive BCG signals. Considering that the limited variability of the inter-beat intervals, the rhythmic confidence representing the similarity among the heartbeat intervals can quantified as the robustness of the detection results.

\subsection{Hybrid Detection Using Both MC and RC}

Considering that the detected results of both TM and DL satisfy the constraints of MC and RC, we propose a hybrid detection scheme by using a comprehensive confidence index integrating MC and RC, which can be written by

\begin{equation}
\label{eq:F1}
F = w_1 c_1 - w_2 c_2
\end{equation}
where $w_1>0$ and $w_2>0$ are the weighting factors of MC index ($c_1$) and RC index ($c_2$), respectively, and the values of $c_1$ and $c_2$ are calculated by (\ref{eq:c1}) and (\ref{eq:c2}), respectively. Recalling (\ref{eq:c1}) and (\ref{eq:c2}), larger value of $F$ corresponds to higher confidence, where the corresponding result of either TM- and DL-based methods should be retained. As a consequence, the hybrid method is expected to enjoy the complementary advantages of both heartbeat interval detection methods, and in turn, to achieve superior performance in comparison with either one.

\section{Numerical results}
\label{S3}

\subsection{Baseline Methods}

\subsubsection{TM-based Heartbeat Detection}
Without the loss of generality, in the process of TM, a dynamic template is modeled by consecutive time-shifted BCG signals, and then used to heartbeat detection. As has been verified by the existing pioneer studies, we refer to \cite{liang2020effective} as a representative since it performs superior to the conventional TM-based schemes. The step-by-step procedures are briefly described as follows.

Step 1: artifact motions are detected and removed from the raw BCG signals.

Step 2: initial detection of BCG signals is performed, by which the template of BCG is modeled and refined with specific criteria \cite{liang2020effective}.

Step 3: the J-peak of each BCG signal is detected in a forward and backward process with the modeled template by using both cross-correlation and dynamic time wrapping.

Step 4: heartbeat detection is obtained with the fused detection of BCG time-series signals.

\subsubsection{DL-based Heartbeat Detection}
As validated by \cite{zhou20211d} and \cite{chen2020ballistocardiography}, specifically designed DL model is able to distinguish the IJK complex among the waveforms of a whole BCG signal period in practical scenarios. Taking the advantage of recurrent neural network for time-series data, we refer to UNet-BiLSTM\cite{mai2022non} as the representative of DL-based model, since it performs well for heartbeat detection of BCG signals among the state-of-the-arts.

Regarding the model design, UNet is a five-layer network structure. The size of the input and output data of the network is 1200 (corresponding to 12 s). A 2$\times$1 max-pooling layer is used for down-sampling. Each down-sampling block contains two convolution layers with kernel size of 17$\times$1. The hidden layers of UNet convolutional layers are 32, 64, 128, 256 and 512, respectively. And each convolutional layer is followed by a rectified linear unit and a batch normalization layer. A BiLSTM with two hidden layers of size 1024 is added to the UNet extended path to further extract temporal features of the BCG signal. The final output result is obtained by inputting the feature vectors into two convolution layers with kernel size of 1$\times$1. The dropout layer is added after the BiLSTM layer and the output layer, respectively. When IJK complex is detected, J-peak of within IJK complex can be obtained correspondingly by finding the local maximum. 

In the experiments, the binary cross-entropy (BCE) loss function and Adam optimizer were used. The dataset was divided into training and testing sets by a 6-fold cross-validation. In each iteration, the BCG of 15 subjects was used for training, and the BCG of other 3 subjects was used for testing. In addition, the data of the rest 16 subjects were used as the hold-out dataset to further test the performance of heartbeat detection. The batch size and the epoch were set to 32 and 30, respectively. Early stopping was implemented to prevent overfitting. The learning rate started at 0.001 and decayed by 20\% every 5 epochs.

For all numerical computing, the hardware devices used in the experiment are Intel (R) Xeon (R) Gold 5118 CPU and NVIDIA GeForce RTX 2080 Ti GPU. The software environment adopted Python 3.8 and used the PyTorch framework to build the neural network model.

\subsection{Evaluation Metrics}
Similar to the pioneer studies \cite{paalasmaa2014adaptive},\cite{wen2019correlation} and \cite{jung2021accurate}, absolute interval error (denoted by $E_\mathrm{abs}$), and precision (denoted by $Pre$) are considered as metrics to assess the performance of BCG heartbeat detection in comparison with ECG (as the latter served as the gold standard in related works). $E_\mathrm{abs}$ refers to the average absolute error between the predicted heartbeat intervals and the true heartbeat intervals, defined as\begin{equation}E_\mathrm{abs}=\frac{1}{N_\mathrm{BCG}}\sum_{i=0}^{N_\mathrm{BCG}-1}|RR(i)-JJ(i)|,\label{eq5}\end{equation}
where $N_\mathrm{BCG}$ denotes the interval numbers of the detected BCG signals, $RR(i)$ and $JJ(i)$ are the $i$ heartbeat interval detected by ECG and BCG signals, respectively. In addition, $Pre$ refers to the accuracy of the detected heartbeats, defined as\begin{equation}Pre=\frac{N_\mathrm{correct}}{N_\mathrm{correct}+N_\mathrm{incorrect}}\times100\%,\label{eq6}\end{equation}
where $N_\mathrm{correct}$ represents the number of intervals with absolute error less than or equal to 30 ms, and $N_\mathrm{incorrect}$ represents the number of intervals with absolute error greater than 30 ms.
\footnote{The signal acquisition device of ECG and BCG may have a ms-level mismatch in sampling frequency due to different crystal oscillators.} In addition, the ratio of coverage between the duration of effectively detected BCG and the total recorded time can also be used as a metric to evaluate the effectiveness of heartbeat detection, defined as:\begin{equation}Coverage=\frac{D_\mathrm{detection}}{D_\mathrm{total}}\times100\%,\label{eq7}\end{equation}
where $D_\mathrm{detection}$ and $D_\mathrm{total}$ are the durations of the detected BCG and the total recording of the raw BCG signals.

\subsection{Confidence Analysis-based Heartbeat Detection}

\begin{table}
\centering
\caption{MC analysis-based heartbeat detection using the representative TM and DL}
\begin{tabular}{cccc}
\hline
Threshold & $E_\mathrm{abs}$ (ms) & $Pre$ (\%)  &  $Coverage$ (\%)\\
\multirow{1}{*}{$T_\mathrm{MC}$} & \multicolumn{1}{c}{TM / DL} & \multicolumn{1}{c}{TM / DL}& \multicolumn{1}{c}{TM / DL} \\  \hline
—— &   50.01 / 30.86 & 79.88 / 85.98 & 92.12 / 91.05 \\
0.55 & 49.84 / 29.89 & 80.05 / 86.58 & 91.45 / 90.34 \\
0.60 & 49.22 / 29.08 & 80.33 / 87.00 & 90.73 / 89.53 \\
0.65 & 47.69 / 27.74 & 80.98 / 87.73 & 89.35 / 88.08 \\
0.70 & 44.58 / 25.78 & 82.24 / 88.76 & 86.96 / 85.87 \\
0.75 & 39.71 / 23.02 & 84.25 / 90.24 & 83.14 / 82.42 \\
0.80 & 33.10 / 19.66 & 87.12 / 92.17 & 77.30 / 76.94 \\
0.85 & 25.04 / 15.65 & 90.92 / 94.62 & 68.45 / 68.36 \\
0.90 & 15.71 / 11.24 & 95.48 / 97.40 & 53.84 / 53.35 \\
\hline
\end{tabular}
\begin{tablenotes}    
\footnotesize              
\item[1]       
\end{tablenotes}           
\label{table1}
\end{table}

\begin{table}
\centering
\caption{RC analysis-based heartbeat detection using the representative TM and DL}
\begin{tabular}{cccc}
\hline
Threshold & $E_\mathrm{abs}$ (ms) & $Pre$ (\%) & $Coverage$ (\%) \\
\multirow{1}{*}{$T_\mathrm{RC}$} & \multicolumn{1}{c}{TM / DL} & \multicolumn{1}{c}{TM / DL}& \multicolumn{1}{c}{TM / DL} \\  \hline
—— &   50.01 / 30.86 & 79.88 / 85.98 & 92.12 / 91.05 \\
0.40 & 46.54 / 29.21 & 80.83 / 86.61 & 89.24 / 89.86 \\
0.35 & 45.75 / 28.25 & 81.08 / 86.86 & 88.32 / 89.30 \\
0.30 & 45.22 / 26.69 & 81.21 / 87.20 & 87.26 / 88.24 \\
0.25 & 42.90 / 24.79 & 81.71 / 87.60 & 86.01 / 87.15 \\
0.20 & 36.51 / 23.24 & 83.40 / 87.96 & 82.79 / 86.38 \\
0.15 & 27.36 / 21.74 & 86.44 / 88.45 & 77.02 / 85.29 \\
0.10 & 17.54 / 18.90 & 91.27 / 89.92 & 67.85 / 81.33 \\
0.05 & 10.15 / 12.17 & 96.40 / 94.81 & 55.70 / 66.18 \\ 
\hline
\end{tabular}
\begin{tablenotes}    
\footnotesize               
\item[1]       
\end{tablenotes}          
\label{table2}
\end{table}

Firstly, we show the performance of heartbeat detection by using the representative TM- and DL-based methods in this paper. As shown in the first row of Table \ref{table1}, the representative DL-based method performs superior to that using TM in terms of $E_\mathrm{abs}$ and $Pre$, respectively, whereas, the representative TM-based method has a slightly higher effective detection ratio in terms of $Coverage$ in (\ref{eq7}), compared to DL. The result is expected, and is consistent to the pioneer study \cite{mai2022non}.

Taking the representative TM- and DL-based methods as the baseline, we examine the feasibility of the proposed MC and RC for heartbeat detection, respectively. 

In order to gain an insight into the confidence analysis in terms of morphology and rhythm, different levels of confidence w.r.t. MC and RC, i.e. $T_\mathrm{MC}$ and $T_\mathrm{RC}$ in (\ref{eq:c1}) and (\ref{eq:c2}) are considered. The numerical results are listed in Tables \ref{table1} and \ref{table2}. As can be seen from Tables \ref{table1} and \ref{table2}, either increasing $T_\mathrm{MC}$ or decreasing $T_\mathrm{RC}$, the performance of TM and DL grows directly proportional to the so designed level of MC and RC. In particular, when $T_\mathrm{MC} = 0.9$ or $T_\mathrm{RC} = 0.05$, the performance of heartbeat detection, using either TM- or DL-based method, is significantly outperforms that without confidence analysis, and almost comparable to the performance of gold standard with ECG monitoring, but at the expense of dropping a relatively large proportion of data with low confidence.

\subsection{Hybrid Detection Using TM and DL}

\begin{table}
\renewcommand\arraystretch{1.2}
\centering
\caption{The hybrid detection using both MC and RC with different weighting factor ratios}
\begin{tabular}{cccc}\hline  
$\omega_1: \omega_2$    & $E_\mathrm{abs}$ (ms) &  $Pre$ (\%)& $Coverage$ (\%) \\ \hline
3:1  & 22.04 & 90.63 & 86.46 \\ 
2:1  & 21.62 & 90.81 & 86.46 \\ 
1:1  & 21.07 & 91.03 & 86.46 \\  
1:2  & 20.78 & 91.13 & 86.47 \\  
1:3  & 20.73 & 91.13 & 86.47 \\ 
1:4  & 20.73 & 91.11 & 86.47 \\  
1:5  & 20.75 & 91.10 & 86.48 \\   \hline
Mean   & 21.10 & 90.99 & 86.47 \\ 
Std  & 0.52 & 0.20 & 0.01 \\  \hline   
\end{tabular}
\begin{tablenotes}    
\footnotesize              
\item[1] We set $T_\mathrm{MC} = 0.75$ and $T_\mathrm{RC} = 0.20$ as an example.       
\end{tablenotes}            
\label{table3}
\end{table}

Next, we examine the performance of the proposed hybrid detection using TM and DL. Recall (\ref{eq:F1}) that the proposed hybrid detection, in principle, is based on the detected results of TM and DL with confidence constraints by MC and RC simultaneously. We consider different weighting factors w.r.t. MC and RC, i.e., $w_1$ and $w_2$, and list the results of hybrid detection as Table \ref{table3}. As can be seen in Table \ref{table3}, the performance of hybrid detection is fairly robust to the weighting factors of MC and RC over a wide range. Therefore, to facilitate the analysis, we consider $w_1:w_2 =  1:3$ in the following experiments. 

\begin{table}
\centering
\caption{The hybrid detection using both MC and RC}
\begin{tabular}{cccc}
\hline
Threshold & \multirow{2}{*}{$E_\mathrm{abs}$ (ms)} &  \multirow{2}{*}{$Pre$ (\%)} &   \multirow{2}{*}{$Coverage$ (\%)} \\
$T_\mathrm{MC}, T_\mathrm{RC}$  &   &   & \\\hline 
0.55, 0.40   &    30.41  &   87.20  &  94.21\\  
0.60, 0.35   &    29.59  &   87.56  &  93.47 \\  
0.65, 0.30   &    28.10  &   88.16  &  92.18 \\  
0.70, 0.25   &    25.05  &   89.34  &  89.96 \\    
0.75, 0.20   &    20.73  &   91.13  &  86.47 \\    
0.80, 0.15   &    16.62  &   93.28  &  81.14 \\    
0.85, 0.10   &    12.12  &   96.12  &  71.61 \\    
0.90, 0.05   &    8.73   &   98.49  &  54.13 \\     
\hline
\end{tabular}
\begin{tablenotes}    
\footnotesize               
\item[1]        
\end{tablenotes}            
\label{table4}
\end{table}

We then evaluate the performance of the hybrid detection using TM and DL. The proposed hybrid detection is to select the result of a higher confidence w.r.t. MC and RC from the potential candidates obtained from TM and DL, and thus, can enjoy the complementary advantages of both TM and DL. As shown in Table \ref{table4}, the proposed hybrid detection performs superior to those of TM- and DL-based methods as shown in Tables \ref{table1} and \ref{table2}. To be specific, with similar performance metric in terms of detection coverage, i.e., 90\%, the proposed hybrid detection (with $T_\mathrm{MC} = 0.70$ and $T_\mathrm{RC} = 0.25$) yields the heartbeat detection accuracy of $E_\mathrm{abs} = 25.05$ ms and $Pre = 89.34\%$, which significantly outperforms that of TM and DL corresponding to $T_\mathrm{MC} = 0.60$ and $T_\mathrm{RC} = 0.40$, respectively. On the other hand, with the similar level of detection accuracy, i.e., the hybrid detection with $E_\mathrm{abs} \approx 20$ ms and $Pre \geq 91\%$ (corresponding to the confidence $T_\mathrm{MC} = 0.75$ and $T_\mathrm{RC} = 0.20$), the coverage of detection can be significantly improved compared to TM and DL. The results above demonstrate the advantages of the hybrid detection.

\subsection{Per-sample Analysis}

\begin{table*}
\centering
\caption{Per-user performance w.r.t. subjects with normal sinus rhythm}
\begin{tabular}{ccccccc}
\hline
\multirow{2}{*}{No.} && \multicolumn{1}{c}{$E_\mathrm{abs}$ (ms)} && \multicolumn{1}{c}{$Pre$ (\%)} &&  \multicolumn{1}{c}{$Coverage$ (\%)} \\ 
 && TM /  DL / \textbf{Proposed} &&   TM /  DL / \textbf{Proposed} &&  TM /  DL / \textbf{Proposed} \\ \hline
1	&&	28.54 	/	15.93 	/	\textbf{14.53}	&&	88.48 	/	93.10 	/	\textbf{96.10}	&&	97.40 	/	99.64 	/	\textbf{97.33}	\\
2	&& \ 6.53 	/ \	5.74 	/ \	\textbf{4.62}	&&	99.15 	/	99.77 	/	\textbf{99.84}	&&	99.51 	/	99.89 	/	\textbf{99.62}	\\
3	&&	37.34 	/	15.01 	/	\textbf{17.30}	&&	81.31 	/	89.54 	/	\textbf{90.53}	&&	97.92 	/	99.61 	/	\textbf{98.41}	\\
4	&&	56.69 	/	55.77 	/	\textbf{35.10}	&&	71.87 	/	68.10 	/	\textbf{80.93}	&&	93.42 	/	92.86 	/	\textbf{81.47}	\\
5	&&	20.68 	/	10.04 	/	\textbf{11.07}	&&	91.57 	/	97.99 	/	\textbf{97.35}	&&	98.86 	/	99.56 	/	\textbf{99.19}	\\
6	&&	13.84 	/ \	9.17 	/ \	\textbf{9.11}	&&	96.68 	/	99.16 	/	\textbf{99.19}	&&	98.78 	/	99.83 	/	\textbf{99.17}	\\
7	&&	11.23 	/	25.02 	/	\textbf{10.41}	&&	95.87 	/	87.89 	/	\textbf{96.64}	&&	98.73 	/	92.50 	/	\textbf{97.04}	\\
8	&&	15.26 	/	11.83 	/ \	\textbf{9.17}	&&	96.08 	/	97.38 	/	\textbf{98.11}	&&	97.40 	/	99.07 	/	\textbf{97.92}	\\
9	&&	51.09 	/	22.85 	/	\textbf{21.90}	&&	73.45 	/	82.96 	/	\textbf{86.04}	&&	95.01 	/	97.64 	/	\textbf{93.57}	\\
10	&&	36.63 	/	22.95 	/	\textbf{17.75}	&&	86.05 	/	88.21 	/	\textbf{92.31}	&&	94.12 	/	99.27 	/	\textbf{93.93}	\\
11	&&	51.71 	/	22.30 	/	\textbf{18.59}	&&	76.40 	/	90.05 	/	\textbf{92.59}	&&	91.28 	/	91.23 	/	\textbf{86.67}	\\
12	&&	115.21 	/	65.16 	/	\textbf{33.10}	&&	58.25 	/	70.33 	/	\textbf{84.92}	&&	76.95 	/	79.60 	/	\textbf{58.80}	\\
13	&&	41.92 	/	30.98 	/	\textbf{22.41}	&&	81.49 	/	90.35 	/	\textbf{92.12}	&&	89.19 	/	62.66 	/	\textbf{79.49}	\\
14	&&	26.93 	/	18.22 	/	\textbf{15.02}	&&	86.73 	/	91.12 	/	\textbf{92.92}	&&	98.59 	/	98.51 	/	\textbf{95.82}	\\
15	&&	14.30 	/ \	8.52 	/ \	\textbf{7.01}	&&	95.03 	/	98.59 	/	\textbf{98.71}	&&	98.58 	/	97.88 	/	\textbf{97.42}	\\
16	&&	90.79 	/	62.75 	/	\textbf{26.07}	&&	64.68 	/	70.29 	/	\textbf{88.79}	&&	87.41 	/	86.41 	/	\textbf{64.64}	\\
17	&&	39.48 	/	28.04 	/	\textbf{18.99}	&&	83.86 	/	89.28 	/	\textbf{92.70}	&&	96.14 	/	95.01 	/	\textbf{88.08}	\\
18	&&	44.68 	/	25.79 	/	\textbf{21.27}	&&	77.17 	/	87.61 	/	\textbf{89.98}	&&	95.20 	/	96.16 	/	\textbf{90.11}	\\
19	&&	41.75 	/	14.15 	/	\textbf{14.47}	&&	82.99 	/	96.66 	/	\textbf{96.47}	&&	97.36 	/	97.39 	/	\textbf{97.26}	\\
20	&&	46.19 	/	21.52 	/	\textbf{13.03}	&&	80.90 	/	90.29 	/	\textbf{95.74}	&&	94.88 	/	94.32 	/	\textbf{87.98}	\\
21	&&	21.01 	/	16.69 	/ \	\textbf{9.61}	&&	94.87 	/	94.79 	/	\textbf{98.20}	&&	90.45 	/	93.22 	/	\textbf{96.07}	\\
22	&&	31.98 	/	13.20 	/	\textbf{13.69}	&&	85.69 	/	96.44 	/	\textbf{95.19}	&&	96.61 	/	91.53 	/	\textbf{95.86}	\\
23	&&	83.48 	/	28.53 	/	\textbf{18.16}	&&	68.95 	/	85.01 	/	\textbf{93.81}	&&	84.71 	/	91.63 	/	\textbf{79.92}	\\\hline  
Mean	&&	40.32 	/	23.92 	/	\textbf{16.63}	&&	83.37 	/	89.34 	/	\textbf{93.44}	&&	94.28 	/	93.71 	/	\textbf{90.25}	\\
Std	&&	26.76 	/	16.36 	/ \	\textbf{7.68}	&&	10.97 	/ \	9.11 	/ \	\textbf{4.89}	&& \	5.51 	/ \	8.40 	/	\textbf{10.99}	\\
\hline  
\end{tabular}
\label{table5}
\end{table*}

\begin{table*}
\centering
\caption{Per-user performance w.r.t. subjects suffering from arrhythmia}
\begin{tabular}{ccccccc}

\hline
\multirow{2}{*}{No.} && \multicolumn{1}{c}{$E_\mathrm{abs}$ (ms)} & \multicolumn{1}{c}{ }& \multicolumn{1}{c}{$Pre$ (\%)} & \multicolumn{1}{c}{ } & \multicolumn{1}{c}{$Coverage$ (\%)} \\
 && TM /  DL / \textbf{Proposed} &  & TM /  DL / \textbf{Proposed} &  & TM /  DL / \textbf{Proposed} \\ \hline
24	&&	62.83 	/	34.00 	/	\textbf{24.85}	&&	71.98 	/	79.57 	/	\textbf{85.96}	&&	88.58 	/	95.26 	/	\textbf{84.18}	\\
25	&&	39.17 	/	32.26 	/	\textbf{20.68}	&&	79.84 	/	81.19 	/	\textbf{90.05}	&&	95.25 	/	98.64 	/	\textbf{92.80}	\\
26	&&	104.96 	/	47.43 	/	\textbf{33.20}	&&	74.55 	/	76.23 	/	\textbf{84.43}	&&	63.03 	/	67.72 	/	\textbf{64.01}	\\
27	&&	94.15 	/	98.17 	/	\textbf{64.30}	&&	60.71 	/	49.54 	/	\textbf{66.62}	&&	76.24 	/	91.12 	/	\textbf{76.12}	\\
28	&&	51.61 	/	65.30 	/	\textbf{27.05}	&&	79.08 	/	77.27 	/	\textbf{89.34}	&&	87.81 	/	65.19 	/	\textbf{72.64}	\\
29	&&	98.52 	/	49.63 	/	\textbf{44.76}	&&	58.37 	/	68.79 	/	\textbf{75.14}	&&	86.98 	/	74.40 	/	\textbf{67.18}	\\
30	&&	29.52 	/	16.87 	/	\textbf{12.08}	&&	89.01 	/	95.35 	/	\textbf{96.01}	&&	96.74 	/	93.52 	/	\textbf{92.86}	\\
31	&&	158.81 	/	71.63 	/	\textbf{25.96}	&&	40.29 	/	74.01 	/	\textbf{89.58}	&&	83.68 	/	70.47 	/	\textbf{39.34}	\\
32	&&	61.32 	/	49.26 	/	\textbf{37.14}	&&	74.33 	/	78.89 	/	\textbf{83.50}	&&	91.74 	/	89.42 	/	\textbf{85.61}	\\
33	&&	40.88 	/	16.29 	/	\textbf{15.23}	&&	82.89 	/	95.28 	/	\textbf{95.06}	&&	96.24 	/	95.68 	/	\textbf{93.23}	\\
34	&&	31.28 	/	18.37 	/	\textbf{17.17}	&&	87.31 	/	92.17 	/	\textbf{93.51}	&&	97.36 	/	98.70 	/	\textbf{96.15}	\\ \hline 
Mean	&&	70.28 	/	45.38 	/	\textbf{29.31}	&&	72.58 	/	78.94 	/	\textbf{86.29}	&&	87.60 	/	85.47 	/	\textbf{78.56}	\\
Std	&&	39.88 	/	25.69 	/	\textbf{15.14}	&&	14.43 	/	13.11 	/ \	\textbf{8.83}	&&	10.39 	/	13.17 	/	\textbf{17.13}	\\

 \hline  
\end{tabular}
\label{table6}
\end{table*}

To gain an insight into the proposed hybrid detection, we provide the results w.r.t. per-sample analysis, where the subjects evolved in the experiments are grouped depending on the rhythm of heartbeat, i.e., sinus rhythm and arrhythmia, as listed in Tables \ref{table5} and \ref{table6}, respectively. As shown in Table \ref{table5}, it can be seen that the hybrid detection performs superior to both TM and DL for all subjects with sinus rhythm, i.e., the detection performance in terms of mean $E_\mathrm{abs}$, in comparison with that of TM and DL, are significantly improved by 23.69 ms and 7.29 ms, corresponding to the performance gain of 58.75\% and 30.47\%, respectively. In addition, the standard deviation of per-sample results in terms of $E_\mathrm{abs}$ and $Pre$ are much smaller than those of TM and DL. This demonstrated that the confidence analysis-based hybrid detection performs robust to different subjects. It is noted that the hybrid detection leads to a relatively large decrease of detection coverage for subject No. 12, i.e. 58.80\%. The potential reason is that the ratio of the motion artifact during the whole-night non-invasive monitoring is relatively high, resulting in a high proportion of low signal quality, and accordingly inaccurate detection performance, which is removed under the confidence analysis of MC and RC. Anyway, the performance of the averaged detection coverage for all 23 subjects can reach 90.25\%, which is comparable to that of TM and DL.

For subjects with arrhythmia, the proposed hybrid detection also achieves significantly higher accuracy than TM and DL. Compared with the experiments evolving subjects of sinus rhythm, the performance of TM, DL, and hybrid detection for subjects with arrhythmia all decreased to some extent, but still, the hybrid detection performs more robust for different subjects. It is worth noting that the mean coverage of the hybrid detection is 78.56\%, which is lower than that of TM and DL. The results are expected, since the evolved subjects with arrhythmia (i.e., No. 27 and No. 31) suffer from severe cardiac disease, which have influenced the cardiac ejection, leading to a high ratio of unreliable BCG recording with low signal quality, and correspondingly, the detected results those were unable to meet the requirement of MC and RC are completely removed during the hybrid detection process. Anyway, from the perspective of per-sample analysis, we further demonstrated the effectiveness of the hybrid detection, which can achieve more robust and flexible performance of heartbeat detection than TM and DL by adjusting the confidence levels of MC and RC.

\subsection{Discussion}
The experimental results above revealed that the hybrid detection exhibits superior performance compared to TM and DL without degradation in detection coverage. The enhanced performance benefits from the ability of the hybrid detection method that can enjoy the complementary advantages of TM and DL.

\begin{table*}
\centering
\caption{Comparisons w.r.t. different levels of signal quality and heart rate variability}
\begin{tabular}{cccccccc}

\hline
\multirow{2}{*}{}   & \multirow{2}{*}{Method}   && \multicolumn{1}{c}{$E_\mathrm{abs}$ (ms)} & \multicolumn{1}{c}{ }& \multicolumn{1}{c}{$Pre$ (\%)} & \multicolumn{1}{c}{ }& \multicolumn{1}{c}{$Coverage$ (\%)} \\ 
\multirow{2}{*}{}  &\multirow{2}{*}{}  && A / B /  C & & A / B /  C & & A / B / C \\ \hline
\multirow{3}{*}{Low HRV}& \multicolumn{1}{c}{TM}    & &  $\underline{8.64}$   /  22.67  /   81.35    & &   $\underline{97.80}$   /  91.13   /  64.30  & &    $\underline{99.08}$  /  97.80  /  $\underline{93.13}$   \\ 
 &  \multicolumn{1}{c}{DL}   &&   9.93   /  $\underline{15.51}$  /   $\underline{40.34}$    & &   97.42  /   $\underline{94.57}$  /  $\underline{79.33}$  & &  98.39   /  $\underline{98.00}$  /  88.88   \\ 
 & \textbf{Proposed}    & &  \textbf{6.54}   /  \textbf{12.11}  /    \textbf{28.20}   & &   \textbf{99.01}   /   \textbf{96.76}  /  \textbf{85.28}  & &   \textbf{99.36}  /  \textbf{98.10}  /   \textbf{75.34}   \\ \hline 
 \multirow{3}{*}{High HRV} & \multicolumn{1}{c}{TM}    &&   $\underline{10.78}$   /  44.69  /    109.28   & &   $\underline{97.08}$   /  81.97   /  55.24  & &  $\underline{98.10}$    /  $\underline{92.45}$  /  $\underline{87.17}$   \\
  & \multicolumn{1}{c}{DL}  &&   14.95   /  $\underline{34.74}$  /   $\underline{62.63}$    & &   95.43   /  $\underline{84.27}$   /  $\underline{67.64}$ & &   95.29  / 91.49  /  80.93 \\ 
 & \textbf{Proposed}   &&   \textbf{8.28}   /  \textbf{23.11}  /   \textbf{42.99}    & &    \textbf{98.16}  /   \textbf{90.33}  /  \textbf{77.07}  & &   \textbf{97.12}  /  \textbf{91.66}   / \textbf{64.22}    \\ 
\hline   
\end{tabular}
\begin{tablenotes}    
\footnotesize               
\item[1] 1. The signal quality is characterized by the morphological similarity between each BCG waveform and the averaged model of BCG over 10 s. Similar to \cite{mai2022non}, the signal quality of BCG signals is divided into 3 levels: Level A, Level B and Level C. Consider space constraint, we do not elaborate on this topic. For more details may refer to \cite{mai2022non}. For the assessment of HRV, we refer to the definition developed by \cite{electrophysiology1996heart}, in which high and low HRV are characterized by the root mean square of successive differences (RMSSD) w.r.t. the heartbeat intervals. 2. The compared results with better performance between TM and DL are underlined.
\end{tablenotes}          
\label{tab1e7}
\end{table*}

\begin{figure*}[tbp]
\centering  
\includegraphics[scale=0.42]{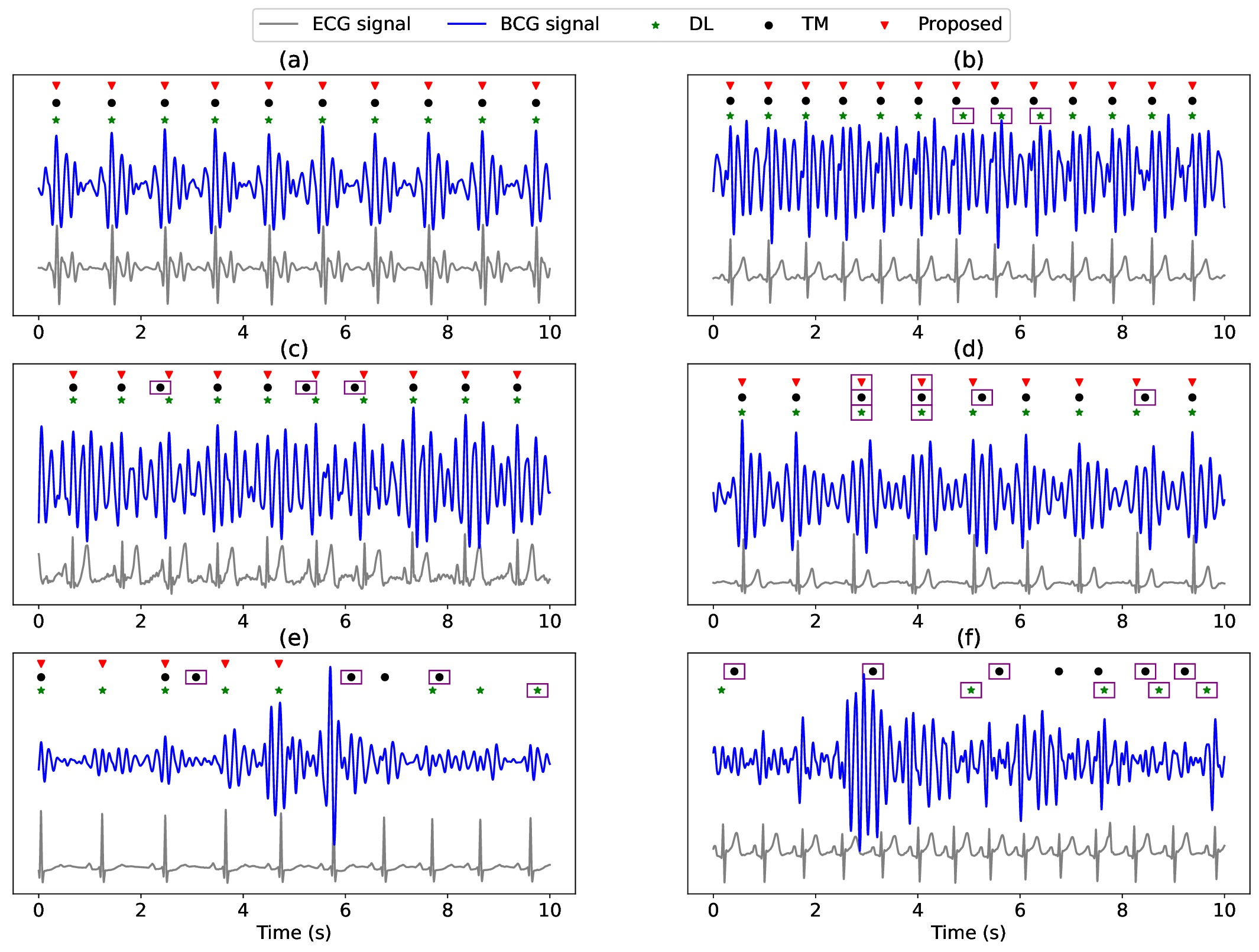} 
\caption{Typical examples of heartbeat detection using the representative TM, DL and the proposed hybrid detection, where boxed results indicate erroneous detections. (a) and (b) are the typical examples for high signal quality of BCG, (c) and (d) are the typical examples for moderate signal quality of BCG, (e) and (f) are the typical examples of low signal quality of BCG, where the waveforms of BCG signals are significantly affected by artifact motion or deep breath during sleep.}
\label{fig4}
\end{figure*}

In this subsection, we further provide a detailed analysis and try to explain how the hybrid detection can achieve the complementary advantages between TM and DL. Referring to the existing studies \cite{mai2022non} and \cite{electrophysiology1996heart}, we categorize the dataset of BCG signals w.r.t. the dimensions of signal quality and rhythm consistency. To be specific, the signal quality of BCG time series is quantified to three levels (Level A, Level B and Level C) w.r.t. the normalized variance across adjacent heartbeats of BCG waveform, which is similar to the definition of SNR \cite{mai2022non}. Recall Fig. \ref{Figure 2}, the BCG signals' morphology of Level A is generally robust and quite identical among consecutive heartbeats, while the signals of Level C are severely disturbed, where the morphological features of BCG could almost disappear. As regards the rhythm consistency, we follow the HRV guideline \cite{electrophysiology1996heart} to quantify BCG signals into two categories using the metric of root mean square value of successive differences (RMSSD), which describes the rhythm consistency between consecutive heartbeat intervals. Generally, RMSSD values of subjects suffering from arrhythmia is much greater than those of sinus rhythm.

As shown in Table \ref{tab1e7}, it can be observed that for BCG signals with different signal qualities and rhythm consistency, either TM or DL has its own advantages for heartbeat detection. To be specific, DL outperforms TM in the cases of moderate and low signal qualities (i.e., Level B and Level C). It is reasonable since DL can benefit from the powerful learning ability that can characterize both the morphological and rhythmic features of BCG signals. On the contrary, for the case of high signal quality (i.e., Level A), TM performs more robust to individual differences w.r.t. the waveform of BCG, and therefore achieve a superior performance to that of DL. The above observations, for the first time, revealed the potential complementarity between TM and DL, and further provide the explanation that how the proposed hybrid detection can enjoy the complementary advantage of TM and DL. As can be observed in Table \ref{tab1e7}, the confidence analysis-based hybrid detection outperforms both TM and DL in different cases w.r.t. signal quality and HRV consistency.

We further present the case studies of heartbeat detection using TM, DL, and the hybrid method under different signal quality and rhythm consistency. The results are shown in Fig. \ref{fig4}. Figs. \ref{fig4}(a) and \ref{fig4}(b) show typical examples of heartbeat detection for high signal quality (characterized by Level A) using the representative TM, DL, and the confidence analysis-based hybrid detection using both TM and DL. Under high signal quality conditions, the TM-based method can extract robust templates that effectively adapt to BCG waveforms with significant individual differences and yield a superior performance to the DL-based method. In this case, the hybrid detection, under the analysis using MC and RC, selects the results of higher confidence. These results are consistent with those listed in Table VII.

In addition, Figs. \ref{fig4}(c) and \ref{fig4}(d) present the typical examples of moderate signal quality (generally characterized by Level B) with normal sinus rhythm and arrhythmia, respectively. Clearly, in moderate signal quality conditions, the IJK complexes of some BCG signals could be interfered with, where some dominant peaks of BCG are replaced by the adjacent waves. As a result, the detection using TM-based method suffers from errors due to the interfered waveforms, as shown in Fig. \ref{fig4}(c). Comparatively, DL exhibits its superiority since the deep model with contextual memory unit can characterize both the rhythmic features and the morphological features of BCG signals, which is consistent with the numerical results listed in Table VII. However, when both the morphology and the rhythm of consecutive BCG are not robust, as shown in Fig. \ref{fig4}(d), i.e., the typical case of moderate signal quality with arrhythmia, neither TM nor DL can perform the correct detection of heartbeat interval due to the interfered signal waveform. In practical conditions, where signal quality and rhythm are not always robust, the hybrid scheme can take advantage of both TM and DL using confidence analysis to preserve detection performance.

We further provide typical examples of heartbeat detection using TM, DL and the hybrid detection-based schemes in practical case, where the signal quality is severely interfered with by motion artifacts, i.e., the amplitudes of motion waveforms are much higher than those of normal BCG signals, as shown in Figs. \ref{fig4}(e) and \ref{fig4}(f). As expected, the DL-based method performs more robust to heartbeat detection in comparison with that using TM. In addition, it is worth noting that the hybrid detection can enjoy the complementary advantages of both TM and DL, while removing the non-robust results of TM and DL through confidence analysis. To be specific, from Fig. \ref{fig4}(f), we can see that the detected results using both TM and DL are all discarded as unreliable based on the statistical confidence analysis w.r.t. MC and RC. Obviously, the discarded results lead to the decrease of detection in terms of signal coverage to some extent, as listed in Table VII, but can significantly improve the performance of detection.

To sum up, from the results shown above, we first demonstrate the advantages of TM and DL in practical conditions, respectively, and then conclude that the proposed hybrid detection can enjoy the complementary advantages of both TM and DL with the constraints of MC and RC, thus it is promising for practical applications of heartbeat detection using BCG signals. The following contents will be addressed in the next stage of the study, 1) evaluating the performance of the hybrid detection by recruiting more samples with severe cardiovascular diseases, such as heart failure, 2) developing more comprehensive confidence constraints.

\section{Conclusion}
\label{S4}
In this paper, we proposed a hybrid detection scheme for BCG signals using both TM and DL. With the confidence analysis w.r.t. the morphology and rhythm using the initially detected BCG signals, the result with a higher confidence can be considered as the hybrid detection, and thereby, enjoys the complementary advantages of TM and DL. Through comprehensive analysis using the practical dataset recorded during nighttime sleep from 34 individuals, we demonstrated that the proposed hybrid scheme outperforms either TM or DL in terms of detection accuracy, and revealed how the hybrid detection
benefits from both TM and DL from different scenarios.

\section*{Acknowledgment}
The illustrative elements in Fig. 2 (e.g., a person, computer, bed) were generated using the Doubao AI tool (by ByteDance). The authors designed the overall layout, integrated these elements, added annotations, and completed the final figure.

\bibliographystyle{IEEEtran}
\bibliography{reference}

\begin{thebibliography}{10}
\providecommand{\url}[1]{#1}
\csname url@samestyle\endcsname
\providecommand{\newblock}{\relax}
\providecommand{\bibinfo}[2]{#2}
\providecommand{\BIBentrySTDinterwordspacing}{\spaceskip=0pt\relax}
\providecommand{\BIBentryALTinterwordstretchfactor}{4}
\providecommand{\BIBentryALTinterwordspacing}{\spaceskip=\fontdimen2\font plus
\BIBentryALTinterwordstretchfactor\fontdimen3\font minus \fontdimen4\font\relax}
\providecommand{\BIBforeignlanguage}[2]{{%
\expandafter\ifx\csname l@#1\endcsname\relax
\typeout{** WARNING: IEEEtran.bst: No hyphenation pattern has been}%
\typeout{** loaded for the language `#1'. Using the pattern for}%
\typeout{** the default language instead.}%
\else
\language=\csname l@#1\endcsname
\fi
#2}}
\providecommand{\BIBdecl}{\relax}
\BIBdecl

\bibitem{rajendra2006heart}
U.~Rajendra~Acharya, K.~Paul~Joseph, N.~Kannathal, C.~M. Lim, and J.~S. Suri, ``Heart rate variability: a review,'' \emph{Med. Biol. Eng. Comput.}, vol.~44, no.~12, pp. 1031--1051, Nov. 2006.

\bibitem{zhu2019heart}
J.~Zhu, L.~Ji, and C.~Liu, ``Heart rate variability monitoring for emotion and disorders of emotion,'' \emph{Physiol. Meas.}, vol.~40, no.~6, p. 064004, Jun. 2019.

\bibitem{shokouhmand2021mean}
A.~Shokouhmand, C.~Yang, N.~D. Aranoff, E.~Driggin, P.~Green, and N.~Tavassolian, ``Mean pressure gradient prediction based on chest angular movements and heart rate variability parameters,'' in \emph{Proc. 43rd Annu. Int. Conf. IEEE Eng. Med. Biol. Soc. (EMBC)}, Mexico, Nov. 2021, pp. 7170--7173.

\bibitem{drew2004practice}
B.~J. Drew, R.~M. Califf, M.~Funk, E.~S. Kaufman, M.~W. Krucoff, M.~M. Laks, P.~W. Macfarlane, C.~Sommargren, S.~Swiryn, and G.~F. Van~Hare, ``{Practice standards for electrocardiographic monitoring in hospital settings: an American Heart Association scientific statement from the Councils on Cardiovascular Nursing, Clinical Cardiology, and Cardiovascular Disease in the Young: endorsed by the International Society of Computerized Electrocardiology and the American Association of Critical-Care Nurses},'' \emph{Circulation}, vol. 110, no.~17, pp. 2721--2746, Oct. 2004.

\bibitem{sandau2017update}
K.~E. Sandau, M.~Funk, A.~Auerbach, G.~W. Barsness, K.~Blum, M.~Cvach, R.~Lampert, J.~L. May, G.~M. McDaniel, M.~V. Perez \emph{et~al.}, ``{Update to practice standards for electrocardiographic monitoring in hospital settings: a scientific statement from the American Heart Association},'' \emph{Circulation}, vol. 136, no.~19, pp. e273--e344, Oct. 2017.

\bibitem{kotalczyk20212020}
A.~Kotalczyk, G.~Y. Lip, and H.~Calkins, ``{The 2020 ESC guidelines on the diagnosis and management of atrial fibrillation},'' \emph{Arrhythmia Electrophysiol. Rev.}, vol.~10, no.~2, pp. 65--67, Jul. 2021.

\bibitem{10288217}
K.~Wang, B.~Feng, D.~Zhao, K.~Xiao, G.~Wu, B.~Shi, R.~Wang, J.~Chen, K.~Li, L.~Li, G.~Zuo, and C.~Shi, ``Noncontact in bed measurements of electrocardiogram using a capacitively coupled electrode array based on flexible circuit board,'' \emph{IEEE Trans. Instrum. Meas.}, vol.~72, pp. 1--14, 2023.

\bibitem{10449345}
L.~Liu, D.~Yu, H.~Lu, C.~Shan, and W.~Wang, ``Camera-based seismocardiogram for heart rate variability monitoring,'' \emph{IEEE J. Biomed. Health. Inf.}, vol.~28, no.~5, pp. 2794--2805, May 2024.

\bibitem{10551856}
B.~Dong, Y.~Liu, K.~Yang, and J.~Cao, ``Realistic pulse waveforms estimation via contrastive learning in remote photoplethysmography,'' \emph{IEEE Trans. Instrum. Meas.}, vol.~73, pp. 1--15, 2024.

\bibitem{liu2021motion}
Y.~Liu, B.~Qin, R.~Li, X.~Li, A.~Huang, H.~Liu, Y.~Lv, and M.~Liu, ``{Motion-robust multimodal heart rate estimation using BCG fused remote-PPG with deep facial ROI tracker and pose constrained Kalman filter},'' \emph{IEEE Trans. Instrum. Meas.}, vol.~70, pp. 1--15, 2021.

\bibitem{10387261}
W.~Lyu, W.~Yuan, J.~Yu, Q.~Wang, S.~Chen, J.~Qin, and C.~Yu, ``Non-contact short-term heart rate variability analysis under paced respiration based on a robust fiber optic sensor system,'' \emph{IEEE Trans. Instrum. Meas.}, vol.~73, pp. 1--13, 2024.

\bibitem{yu2025proof}
B.~Yu, Y.~Chen, D.~Cai, and H.~Zhang, ``A proof-of-concept study on non-contact {BCG}-based cardiac monitoring for in-patients with sleep apnea syndrome using piezoelectric ceramics,'' \emph{IEEE Trans. Instrum. Meas.}, vol.~74, pp. 1--10, 2025.

\bibitem{9103629}
N.~Mora, F.~Cocconcelli, G.~Matrella, and P.~Ciampolini, ``Accurate heartbeat detection on ballistocardiogram accelerometric traces,'' \emph{IEEE Trans. Instrum. Meas.}, vol.~69, no.~11, pp. 9000--9009, Nov. 2020.

\bibitem{escobedo2024bed}
P.~Escobedo, A.~Pousibet-Garrido, N.~L{\'o}pez-Ruiz, M.~A. Carvajal, A.~J. Palma, and A.~Mart{\'\i}nez-Olmos, ``{Bed-based ballistocardiography system using flexible RFID sensors for noninvasive single- and dual-subject vital signs monitoring},'' \emph{IEEE Trans. Instrum. Meas.}, vol.~73, pp. 1--12, 2024.

\bibitem{10835744}
C.~Wu, Q.~Zhang, and G.~Shen, ``Toward robust person identification using {BCG} signals: A multistage fingerprinting approach,'' \emph{IEEE Trans. Instrum. Meas.}, vol.~74, pp. 1--15, 2025.

\bibitem{jiang2022topological}
F.~Jiang, B.~Xu, Z.~Zhu, and B.~Zhang, ``Topological data analysis approach to extract the persistent homology features of ballistocardiogram signal in unobstructive atrial fibrillation detection,'' \emph{IEEE Sens. J.}, vol.~22, no.~7, pp. 6920--6930, Apr. 2022.

\bibitem{su2022atrial}
Q.~Su, Y.~Huang, X.~Wu, B.~Zhang, P.~Lu, and T.~Lyu, ``{Atrial fibrillation detection based on a residual CNN using BCG signals},'' \emph{Electronics}, vol.~11, no.~18, p. 2974, Sep. 2022.

\bibitem{11018225}
Y.~Zhao, Y.~Zheng, C.~Guo, Y.~Huang, B.~Zhang, P.~Lu, T.~Lyu, and X.~Wu, ``Automatic atrial fibrillation detection based on {BCG} using signal transformation and random kernels convolution,'' \emph{IEEE Sens. J.}, vol.~25, no.~14, pp. 26\,944--26\,955, Jul. 2025.

\bibitem{aydemir2019classification}
V.~B. Aydemir, S.~Nagesh, M.~M.~H. Shandhi, J.~Fan, L.~Klein, M.~Etemadi, J.~A. Heller, O.~T. Inan, and J.~M. Rehg, ``Classification of decompensated heart failure from clinical and home ballistocardiography,'' \emph{IEEE Trans. Biomed. Eng.}, vol.~67, no.~5, pp. 1303--1313, May 2019.

\bibitem{feng2023machine}
S.~Feng, X.~Wu, A.~Bao, G.~Lin, P.~Sun, H.~Cen, S.~Chen, Y.~Liu, W.~He, Z.~Pang \emph{et~al.}, ``{Machine learning-aided detection of heart failure (LVEF$\leq$ 49\%) by using ballistocardiography and respiratory effort signals},'' \emph{Front. Physiol.}, vol.~13, p. 1068824, Jan. 2023.

\bibitem{zhan2024non}
J.~Zhan, X.~Wu, X.~Fu, C.~Li, K.-Q. Deng, Q.~Wei, C.~Zhang, T.~Zhao, C.~Li, L.~Huang \emph{et~al.}, ``Non-contact assessment of cardiac physiology using {FO-MVSS}-based ballistocardiography: a promising approach for heart failure evaluation,'' \emph{Sci. Rep.}, vol.~14, no.~1, p. 3269, Feb. 2024.

\bibitem{shin2008automatic}
J.~Shin, B.~Choi, Y.~Lim, D.~Jeong, and K.~Park, ``{Automatic ballistocardiogram (BCG) beat detection using a template matching approach},'' in \emph{Proc. 30th Annu. Int. Conf. IEEE Eng. Med. Biol. Soc. (EMBC)}, Vancouver, BC, Canada, Aug. 2008, pp. 1144--1146.

\bibitem{nagura2018estimation}
M.~Nagura, Y.~Mitsukura, T.~Kishimoto, and M.~Mimura, ``An estimation of heart rate variability from ballistocardiogram measured with bed leg sensors,'' in \emph{Proc. 19th IEEE Int. Conf. Ind. Technol. (ICIT)}, Lyon, France, Feb. 2018, pp. 2005--2009.

\bibitem{paalasmaa2014adaptive}
J.~Paalasmaa, H.~Toivonen, and M.~Partinen, ``Adaptive heartbeat modeling for beat-to-beat heart rate measurement in ballistocardiograms,'' \emph{IEEE J. Biomed. Health. Inf.}, vol.~19, no.~6, pp. 1945--1952, Nov. 2015.

\bibitem{xie2019personalized}
Q.~Xie, M.~Wang, Y.~Zhao, Z.~He, Y.~Li, G.~Wang, and Y.~Lian, ``A personalized beat-to-beat heart rate detection system from ballistocardiogram for smart home applications,'' \emph{IEEE Trans. Biomed. Circuits Syst.}, vol.~13, no.~6, pp. 1593--1602, Dec. 2019.

\bibitem{jiao2021non}
C.~Jiao, C.~Chen, S.~Gou, D.~Hai, B.-Y. Su, M.~Skubic, L.~Jiao, A.~Zare, and K.~Ho, ``Non-invasive heart rate estimation from ballistocardiograms using bidirectional {LSTM} regression,'' \emph{IEEE J. Biomed. Health. Inf.}, vol.~25, no.~9, pp. 3396--3407, Sep. 2021.

\bibitem{mai2022non}
Y.~Mai, Z.~Chen, B.~Yu, Y.~Li, Z.~Pang, and Z.~Han, ``{Non-contact heartbeat detection based on ballistocardiogram using UNet and bidirectional long short-term memory},'' \emph{IEEE J. Biomed. Health. Inf.}, vol.~26, no.~8, pp. 3720--3730, Aug. 2022.

\bibitem{10810416}
C.~Jiao, A.~Yang, H.~Zhao, R.~Yi, S.~Gou, Y.~Sha, W.~Wen, L.~Jiao, and M.~Skubic, ``Self-supervised, non-contact heartbeat detection based on ballistocardiograms utilizing physiological information guidance,'' \emph{IEEE J. Biomed. Health. Inf.}, vol.~29, no.~4, pp. 2589--2602, Apr. 2025.

\bibitem{hai2020heartbeat}
D.~Hai, C.~Chen, R.~Yi, S.~Gou, B.~Y. Su, C.~Jiao, and M.~Skubic, ``Heartbeat detection and rate estimation from ballistocardiograms using the gated recurrent unit network,'' in \emph{Proc. 42nd Annu. Int. Conf. IEEE Eng. Med. Biol. Soc. (EMBC)}, Montreal, QC, Canada, Jul. 2020, pp. 451--454.

\bibitem{6862843}
C.~Orphanidou, T.~Bonnici, P.~Charlton, D.~Clifton, D.~Vallance, and L.~Tarassenko, ``Signal-quality indices for the electrocardiogram and photoplethysmogram: derivation and applications to wireless monitoring,'' \emph{IEEE J. Biomed. Health. Inf.}, vol.~19, no.~3, pp. 832--838, May 2015.

\bibitem{hayn2012qrs}
D.~Hayn, B.~Jammerbund, and G.~Schreier, ``{QRS detection based ECG quality assessment},'' \emph{Physiol. Meas.}, vol.~33, no.~9, p. 1449, Aug. 2012.

\bibitem{8253824}
P.~Mohapatra, P.~Sreeletha~Premkumar, and M.~Sivaprakasam, ``A yellow–orange wavelength-based short-term heart rate variability measurement scheme for wrist-based wearables,'' \emph{IEEE Trans. Instrum. Meas.}, vol.~67, no.~5, pp. 1091--1101, May 2018.

\bibitem{9452120}
P.~Sharma and E.~Rodriguez-Villegas, ``Acoustic sensing as a novel wearable approach for heart rate variability monitoring at the wrist,'' \emph{IEEE Trans. Instrum. Meas.}, vol.~70, pp. 1--12, 2021.

\bibitem{liang2020effective}
J.~Liang, J.~Huang, L.~Mu, B.~Yu, P.~Chen, Z.~Pang, R.~Nie, and H.~Zhang, ``An effective algorithm for beat-to-beat heart rate monitoring from ballistocardiograms,'' \emph{J. Med. Imaging Health Inf.}, vol.~10, no.~3, pp. 633--640, Mar. 2020.

\bibitem{zhou20211d}
T.~Zhou, S.~Men, J.~Liang, B.~Yu, H.~Zhang, and X.~Luo, ``{1D U-net++: an effective method for ballistocardiogram J-peak detection},'' \emph{J. Mech. Med. Biol.}, vol.~21, no.~10, p. 2140058, 2021.

\bibitem{chen2020ballistocardiography}
S.~Chen, F.~Tan, W.~Lyu, and C.~Yu, ``Ballistocardiography monitoring system based on optical fiber interferometer aided with heartbeat segmentation algorithm,'' \emph{Biomed. Opt. Express}, vol.~11, no.~10, pp. 5458--5469, 2020.

\bibitem{wen2019correlation}
X.~Wen, Y.~Huang, X.~Wu, and B.~Zhang, ``A correlation-based algorithm for beat-to-beat heart rate estimation from ballistocardiograms,'' in \emph{Proc. 41st Annu. Int. Conf. IEEE Eng. Med. Biol. Soc. (EMBC)}, Berlin, Germany, Jul. 2019, pp. 6355--6358.

\bibitem{jung2021accurate}
H.~Jung, J.~P. Kimball, T.~Receveur, E.~D. Agdeppa, and O.~T. Inan, ``Accurate ballistocardiogram based heart rate estimation using an array of load cells in a hospital bed,'' \emph{IEEE J. Biomed. Health. Inf.}, vol.~25, no.~9, pp. 3373--3383, Sep. 2021.

\bibitem{electrophysiology1996heart}
T.~F. o. t. E. S. o. C. t. N. A. S. o.~P. Electrophysiology, ``Heart rate variability: standards of measurement, physiological interpretation, and clinical use,'' \emph{Circulation}, vol.~93, no.~5, pp. 1043--1065, Mar. 1996.

\end{thebibliography}

\vfill

\end{document}